# Quantum linear magnetoresistance in NbTe$_2$

Hongxiang Chen[1,2], Zhilin Li[1,2], Xiao Fan[1,2], Liwei Guo[1,2(a)] and Xiaolong Chen[1,2,3(b)]

[1] *Research & Development Center for Functional Crystals, Beijing National Laboratory for Condensed Matter Physics, Institute of Physics, Chinese Academy of Sciences, Beijing 100190, China*
[2] *School of Physical Sciences, University of Chinese Academy of Sciences, Beijing 101408, China*
[3] *Collaborative Innovation Center of Quantum Matter, Beijing 100084, China*

**Abstract** – NbTe$_2$ crystal is quasi-2D layered semimetal with charge density wave ground state showing a distorted-1T structure at room temperature. Here we report the anisotropic magneto-transport properties of NbTe$_2$. An anomalous linear magnetoresistance up to 30% at 3 K in 9 T was observed, which can be well explained by quantum linear magnetoresistance model. Our results reveal that a large quasi-2D Fermi surface and small Fermi pockets with linearly dispersive bands coexist in NbTe$_2$. The comparison with the isostructural material TaTe$_2$ provides more information about the electronic structure evolution with charge density wave transitions in NbTe$_2$ and TaTe$_2$.

**Introduction.** – Transition metal dichalcogenides (TMDCs), MX$_2$ (M=transition metal, X=S, Se, or Te), are a group of layered compounds with weak interlayer interactions governed by van der Waals forces. Depending on the combination of M and X, these dichalcogenides can vary from semiconducting to metallic and even superconducting[1]. The diversity in properties and their thickness sensitivity to [2, 3] have drawn much attention recently. Since Te atom is less negative than S and Se atoms in electronegativity, so more charge is expected to transfer[4] from p bands of Te to the d bands of metal M, leading to an oxidation state M$^{(4-\varepsilon)+}$ in MTe$_2$. The charge-transfer results in the structural distortion and then the unique properties ensue. Recently, considerable attention has been paid on the intrinsic properties of MTe$_2$ and many fascinating phenomena like new Fermions[5], extreme large magnetoresistance(MR)[6], linear MR[7], and superconductivity[8, 9] were found in WTe$_2$, TaTe$_2$, doped or intercalated IrTe$_2$ and so on.

We have previously studied the magneto-transport and magnetic properties of the low temperature phase of TaTe$_2$ (LT-TaTe$_2$) [7]. Soon after our work, the superconductivity with $T_c$~4 K under high pressures in TaTe$_2$ was found[10]. TaTe$_2$ and NbTe$_2$ are isostructural and both show double-zigzag chain-like metal atoms modulations at room temperature[11]. The double-zigzag chain-like structure was observed in the surface morphology studied by transmission electron microscopy (TEM)[12, 13], scanning electron microscopy (STM)[14], and low energy electron diffraction (LEED)[14, 15]. Besides, an intriguing metastable phase under heat pulse[16] or electron beam irradiation of certain intensity[15] has been detected. Similar to NbSe$_2$[17] and doped or intercalated IrTe$_2$, NbTe$_2$ shows a coexistence of charge density wave (CDW) and superconductivity[18, 19].

The superconductivity was found below 0.74 K in NbTe$_2$. Interestingly, it is found the low critical field $H_{c1}$ is extremely small, less than 0.5 Oe. Unlike TaTe$_2$, NbTe$_2$ does not show a further clustering with temperature decreases as appeared in LT-TaTe$_2$. Instead, it exists a significant structure transition from distorted-1T(1T') phase to 1T phase with temperature rising above 550 K (as shown in fig. 1(a))[20]. Hence, NbTe$_2$ is a better platform to study the structural and electronic properties of 1T' phase.

The electronic band structures and Fermi surface of 1T'-NbTe$_2$ have been studied by first-principle calculations and angle-resolved photoelectron emission spectroscopy (ARPES)[14]. The detected Fermi surface is diffusive and an experimental study on the magneto-transport properties in NbTe$_2$ is still lacking. Here, a systematical study on the anisotropic magneto-transport properties of 1T'-NbTe$_2$ is presented, and a detailed comparison in anisotropic magneto-transport properties between NbTe$_2$ and TaTe$_2$ is given to clarify the origin of the quantum linear MR in both materials. Our results on magneto-transport properties of NbTe$_2$ are helpful to reveal the electronic structure evolution with CDW transitions in NbTe$_2$ and TaTe$_2$.

**Experimental.** – Single crystals of NbTe$_2$ were grown by an improved chemical vapor transport (CVT) method[21, 22]. Niobium foil (99.99%), Te powder (99.99%), and little amount of iodine (99%) were loaded into an evacuated silica tube in cross section diameter of 1cm and long 15cm. The tube was heated in a two-zone furnace at 550°C for 1 day, and then rise the temperature to 850°C at the hot end and 750 °C at the cool end. Maintaining this temperature gradient for one week and then cooled naturally, crystals with metal luster were obtained (See the inset of fig. 2(b)) at the cool end. The crystals are usually thin platelets oriented along [001] or long

(a)E-mail: lwguo@iphy.ac.cn
(b)E-mail: chenx29@iphy.ac.cn

ribbons, over half centimeters in at least one dimension. The composition of the prepared NbTe$_2$ sample was checked by energy-dispersive x-ray analysis (EDX). X-ray diffraction (XRD) pattern was collected using a PANalytical X'Pert PRO diffractometer with Cu radiation.

The resistivity measurements were performed on a Quantum Design physical property measurement system (PPMS) with highest magnetic field of 9 T. A standard four probes technique is applied to measure the in-plane resistivity and the MR defined as $(\rho_{xx}(B,T) - \rho_{xx}(0,T))/\rho_{xx}(0,T)$. Field and temperature dependent Hall resistance $R_{xy}(B, T)$ measurements were carried out using a six-probes method. Angle-dependent MR (ADMR) was measured using a rotation option of PPMS.

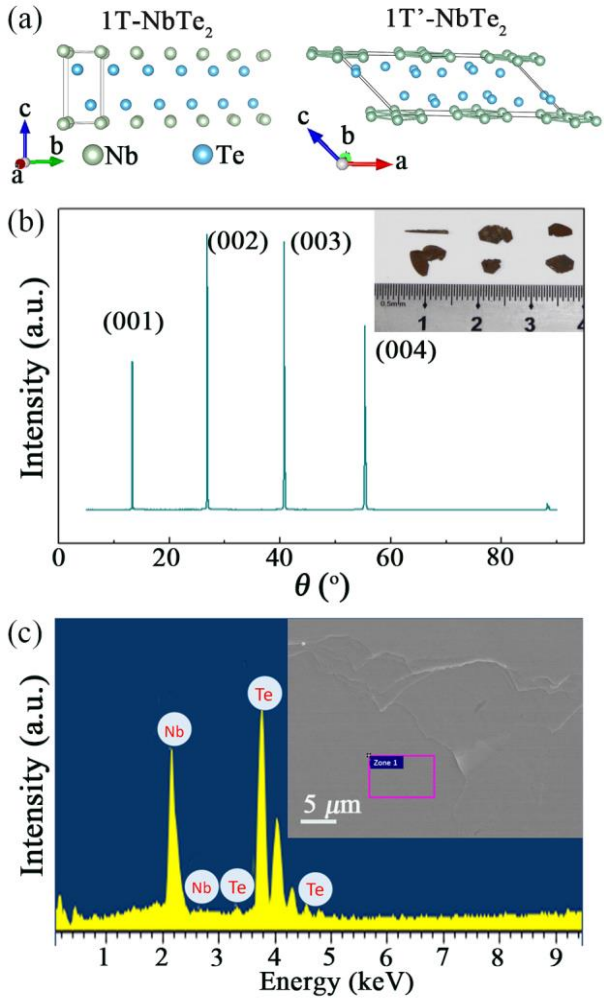

Fig. 1: (Colour on-line) (a) Schematical crystal structures of 1T-NbTe$_2$ and 1T'-NbTe$_2$. The solid lines indicate the unit cells. (b) X-ray diffraction pattern of a TaTe$_2$ single crystal measured at room temperature. Inset: an optical image of typical single crystals from different batches. (c) The energy spectrum data of a cleaved sample surface. Inset: SEM morphology of a cleaved NbTe$_2$ sample surface. The square indicates the analyzed zone of EDX analysis on a sample surface.

**XRD and EDX analysis of NbTe$_2$.** The single crystal samples were checked by XRD diffraction (fig.1(b)). The strong diffraction peaks can be indexed as (00$l$), consistent with the ICDD-PDF 21-0605. Fig. 1(a) presents the crystal structure of 1T'-NbTe$_2$ (Space group C2/m (No. 12), Z=6, a=19.39 Å, b=3.642 Å, c=9.375 Å, $\beta$=134.58 °)[11]. The single crystals are brittle and can be easily cleaved along (001) plane. The layered characteristics can be found in the SEM image showing thin platelet, which means that NbTe$_2$ can be cleaved to mono-layer or few-layers. The cleaved plane is flat and clean without any iodine trace left as supported by EDX data of fig. 1(c). The EDX gave the composition of Nb:Te=35:65, in well agreement with the ratio of chemical stoichiometry of NbTe$_2$. According to the experimental data of XRD and the EDX analysis, it is confirmed that the crystals are NbTe$_2$.

**Magneto-transport Properties of 1T'-NbTe$_2$.** – We measured the temperature-dependent resistivity of several samples from different batches and found the residual resistance ratio ($RRR = \rho_{xx}(300\text{ K})/\rho_{xx}(3\text{ K})$) values are in the range of 10~20. We chose the sample with RRR=20 to study of its magneto-transport properties in detail. Its resistivity is 115.5 $\mu\Omega\cdot$cm at 300 K, and 5.8 $\mu\Omega\cdot$cm at 3 K. The in-plane resistivity $\rho_{xx}$ (0, $T$) (fig. 2(a)) shows a monotonically increasing with temperature, which is a typical metallic behaviour. A small thermal hysteresis is observed near room temperature between the cooling and warming cycles with a rate of 5 K/min. Hall resistances at different temperatures are shown in fig. 2(b). All the Hall resistances show a positive linear dependence on magnetic field $B$, which indicates that almost no carrier concentration changed with increasing field and the dominant charge carriers in NbTe$_2$ are holes, similar to that in TaTe$_2$. While, the Hall coefficient $R_H=d\rho_{xy}(B)/dB$ increases with increasing temperature which is opposite to LT-TaTe$_2$. The net carrier concentration is given by $(n_h - n_e) = 1/(eR_H)$ and the carrier mobility is given by $\mu = R_H/\rho_{xx}$. The carrier concentration ($n_h$-$n_e$) and mobility both decrease with increasing temperature as shown in fig. 2(c). The net carrier concentration ($n_h$-$n_e$) is about $1.7 \times 10^{21}$ cm$^{-3}$ and the mobility is about 600 cm$^2$/(V s) at 3 K.

The transverse MRs versus magnetic field at low temperatures are presented in fig. 3(a). A relatively large MR reached 30% was observed under a field 9 T at 3 K and does not show any signs of saturation in the measured field range. A crossover of MRs from a semi-classical weak-field $B^2$ dependence to a nearly linear $B$ dependence was observed when the magnetic field is beyond a critical field $B^*$. In order to determine the critical field $B^*$, the differential MR versus field, $d$MR/$dB$, is plotted as shown in fig. 3(b). $d$MR/$dB$ is linearly proportional to $B$ with a large positive slope in low fields, but a much reduced slope over a critical field $B^*$ ($B^*$~1.85 T at 3 K), which is determined by an intersection of two linear fitting lines (as shown by the guide lines in fig. 3(b)). In addition, it is noted the critical field shifts to a higher field with increasing temperature, as shown in fig. 3(c). The dependence of the critical field $B^*$ on temperature is shown with blue square in fig. 3(c), which can be well fitted with the red solid curve satisfying $B^* = (1/2e\hbar v_F^2)(E_F + k_BT)^2$ with $v_F \approx 5.65\times10^4$ m/s and $E_F \approx 8.17$ meV. The critical field dependence on temperature suggested the existence of electronic band with linear dispersion[23] in 1T'-NbTe$_2$ and will be discussed in the following.

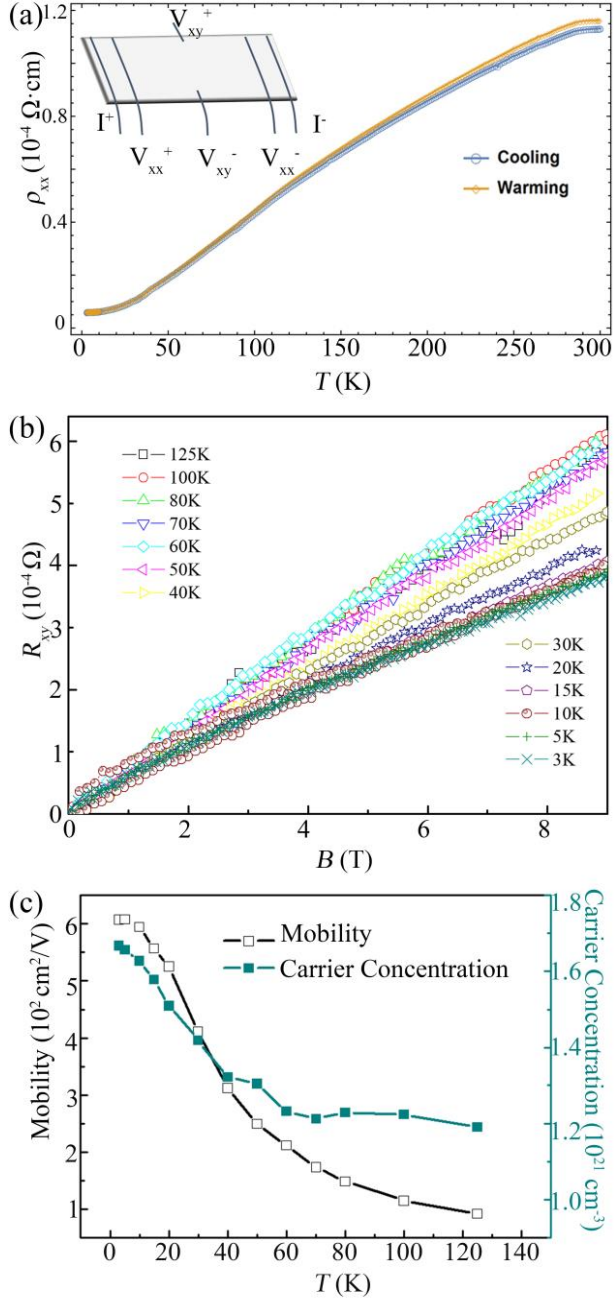

Fig. 2: (Colour on-line) (a) Temperature dependent resistivity of 1T'-NbTe$_2$. Inset: the configuration of six-probes method. (b) The field dependent Hall resistance ($R_{xy}$) at different temperatures. (c) Temperature dependence of carrier concentration and mobility.

From the Hall resistance measurements, the carrier concentration of NbTe$_2$ can be deduced to be in the order of $10^{20}\sim10^{21}$ cm$^{-3}$, which is much higher than the quantum limit to observe a quantum linear MR[24], similar to the case in LT-TaTe$_2$[7]. Therefore, here we discuss the linear MR of NbTe$_2$ with the model of quantum linear MR in layered compounds proposed by Abrikosov[25]. The large carrier concentration in quasi-2D materials indicates the existence of relatively large normal Fermi surface in 1T'-NbTe$_2$. According to the quantum model of Abrikosov[25], the linear MR in 1T'-NbTe$_2$ suggest that apart from the large Fermi surface exists in 1T'-NbTe$_2$, there should exist some small Fermi pockets with almost linear band dispersion in NbTe$_2$. As well known, an applied magnetic field will induce the Landau level (LL) splitting. In the quantum limit at a specific temperature and a field, LL spacing becomes larger than both the Fermi energy $E_F$ and the thermal fluctuations $k_BT$. Consequently, only the lowest Landau level is occupied for the small Fermi pockets and quantum linear MR ensues[24].

For energy band with linear dispersion[26], the splitting between the lowest and the first Landau level can be described by $\Delta_1 = |E_{\pm1} - E_0| = \pm v_F\sqrt{2\hbar eB}$, where $v_F$ is the Fermi velocity. When the quantum limit is approached in a linear energy band, the critical field $B^*$ should satisfy the equation $B^* = (1/2e\hbar v_F^2)(E_F + k_BT)^2$ [23]. As shown in fig. 3(c), the critical field deduced from our 1T'-NbTe$_2$ sample could be well fitted by the equation $B^* = (1/2e\hbar v_F^2)(E_F + k_BT)^2$ [23]. The fitting results (See fig. 3(d)) give the Fermi level $E_F \approx 8.17$ meV, and the Fermi velocity $v_F \approx 5.65\times10^4$ m/s. They are about two times larger and 65.2% higher than the values in LT-TaTe$_2$, respectively. Furthermore, it is deduced that the $\Delta_1=|E_F+k_BT|$ at the critical field $B^*$, satisfying the regime of quantum limit. These results strongly support that the linear MR behaviour originates from the Dirac states formed by the Fermi surface reconstruction in 1T'-NbTe$_2$.

**ADMR of 1T'-NbTe$_2$.** – The ADMR at azimuth angle $\varphi=90°$ are shown in fig. 4(a) and (b). When $\varphi=90°$, the magnetic field was perpendicular to the flowing current along the $b$ axis. The detailed geometry of ADMR measurements, polar angle $\theta$ and azimuth angle $\varphi$ are defined in the inset of fig. 4(a), where a field dependent MR at several angles of $\theta=0°$, 30°, 60°, and 90° are measured. It is noted a linear MR is observed at each polar angle at high fields. Furthermore, the ADMR at 5 K in 9 T is shown in fig. 4(b). It is seen that the ADMR has maximums (MR=30%) when $\theta=0°$ and 180°. The MR decreases to minimums (MR=17%) at $\theta=90°$ and 270°, which decreases by about 43% compared to the maximum. The ADMR could be well fitted by a function of $|\cos\theta|$ as shown in fig. 4(b), which indicates a quasi-2D Fermi surface[27] exists in 1T'-NbTe$_2$. But the slight deviation implies existence of three dimensional electronic transport in 1T'-NbTe$_2$ as reported in the similar material[28].

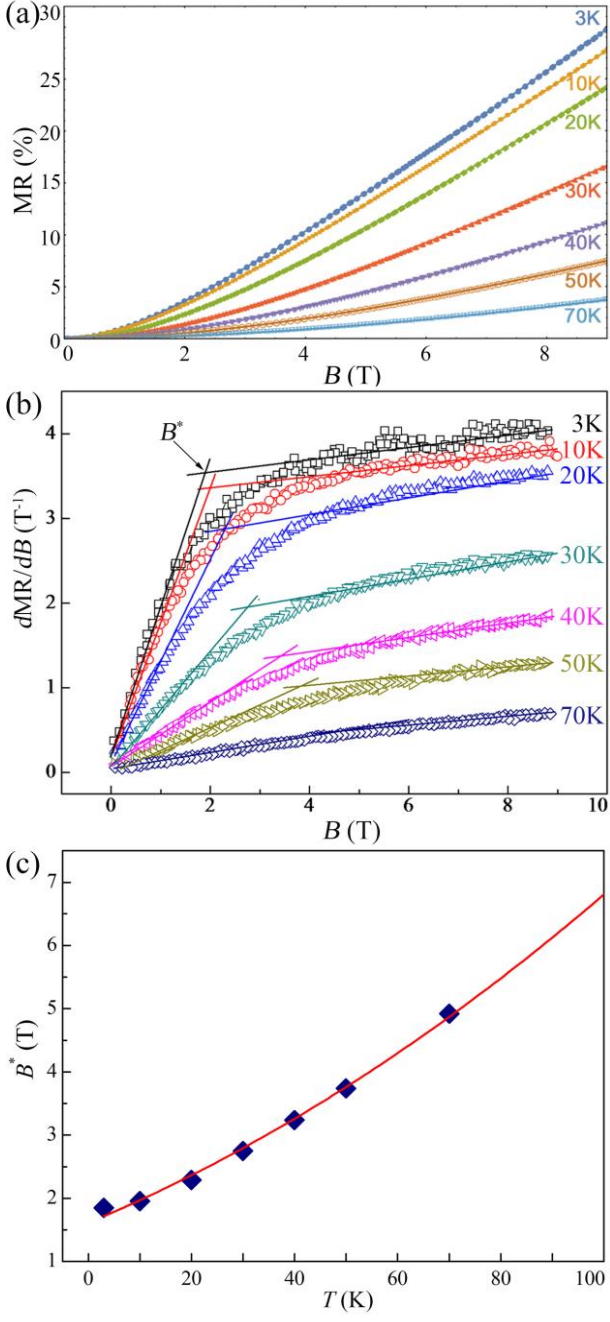

Fig. 3: (Colour on-line) (a) Magnetic field dependent MR at different temperatures. (b) Dependence of differential MR on magnetic field at different temperatures. A critical field $B^*$ is defined as the intersection between the two fitting lines t for $d$MR/$dB$ against $B$ in low field semi-classical regime and high field quantum regime. (c) Dependence of critical field $B^*$ on temperature, data are marked by blue squares.

**Discussion.** – Linear MR under high field seems widely exist in layered CDW and spin density wave (SDW) compounds, such as iron pnictides[23], 2H-NbSe$_2$, 2H-TaSe$_2$[29], LT-TaTe$_2$ and so on. The MRs in these materials are relatively large but smaller than that in some Dirac semimetals[30] or compensated semimetals[6]. The linear MR in 1T'-NbTe$_2$ is quite similar to that of LT-TaTe$_2$. At room temperature, NbTe$_2$ and TaTe$_2$ both are 1T' structure, while TaTe$_2$ shows a further clustering along b axis below 170 K, which is absent in NbTe$_2$. While 1T'-NbTe$_2$ transforms into the 1T phase when temperature is higher than 550 K. As we all know, the CDW transition is accompanied by complicated band-folding[31], and may induced a band with linear dispersion. Linear MRs are observed in both of 1T'-NbTe$_2$ and LT-TaTe$_2$, and can be well explained by the quantum linear MR model. So we suggest that the linear MR in 1T'-NbTe$_2$ and LT-TaTe$_2$ should be resulted from the Fermi pockets with linearly dispersive bands in the 1T' phase. In our previous LT-TaTe$_2$ results, MR shows complicated anisotropy with a two-fold symmetry at low temperatures and high fields which revealed the anisotropy of the small Fermi pockets in Dirac states, while here a simple two-fold symmetric anisotropy was found in 1T'-NbTe$_2$ indicating there exists relatively simple small Fermi pockets with linearly dispersive bands.

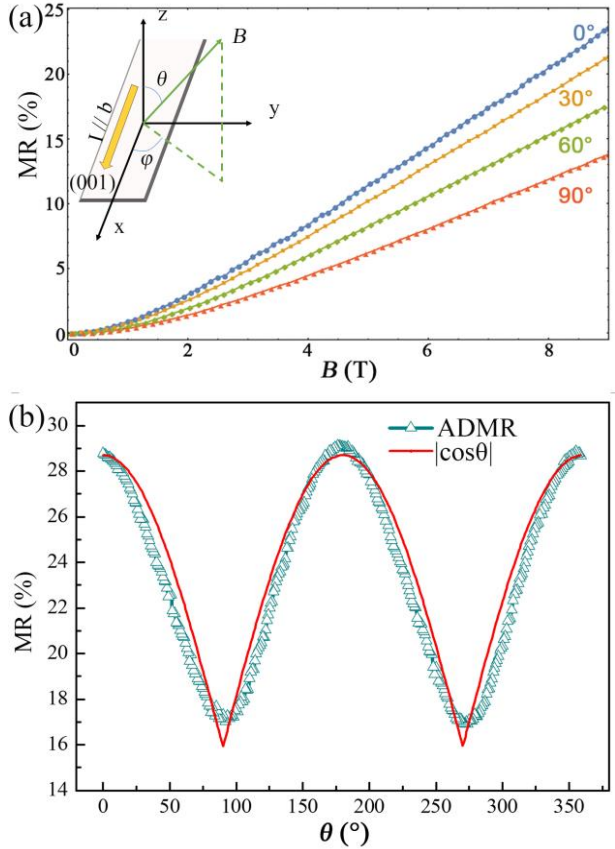

Fig. 4: (Colour on-line) Anisotropic MRs (a) Polar angle $\theta$ ($\varphi$=90°) dependent MR at 5 K in 9 T. The fitting curve fitted by the function of $|\cos\theta|$ is shown by the red curve. (b) Field

dependent MR at four $\theta$ angles ($\theta$=0°, 30°, 60°, and 90°). Inset: the geometry of angle dependent MR measurements.

**CONCLUSION.** – In conclusion, anisotropic magneto-transport properties of 1T'-NbTe$_2$ were studied systematically based on high quality single crystals of NbTe$_2$ with RRR=20. Carrier concentration and mobility of NbTe$_2$ both decreases with increasing temperature. The high carrier concentration around 10$^{21}$ cm$^{-3}$ indicates the existence of relatively large normal Fermi surface. The observed anisotropic linear MR at low temperatures under high fields suggests existing a quasi-2D characteristic in Fermi surface of 1T'-NbTe$_2$. The linear MR and the well fitted critical field $B^*$ versus temperature strongly support the inference of existing small Fermi pockets with linearly dispersive bands in 1T'-NbTe$_2$. The simple two-fold symmetry MR in 1T'-NbTe$_2$ indicates the small linear dispersion Fermi band in 1T'-NbTe$_2$ possessing relative high symmetry compared with that in LT-TaTe$_2$. Our results provide more information about the electronic structure evolutions among 1T, 1T', and LT-TaTe$_2$ phases.

***

The authors thank Mr. J. Zhang of Institute of Physics, Chinese Academy of Sciences, for helpful discussion. This work is financially supported by the National Natural Science Foundation of China (Grant Nos. 91422303, 51532010, 51472265 and 51272279).